\documentstyle[twoside,fleqn,espcrc2,epsf]{article}

\newcommand{\be}{\begin{equation}}
\newcommand{\ee}{\end{equation}}

\newcommand{\AmS}{{\protect\the\textfont2
  A\kern-.1667em\lower.5ex\hbox{M}\kern-.125emS}}

\title{ $B_c$ and $\Upsilon$ Spectra from Lattice NRQCD - Results at $\beta$ = 5.7}

\author{C.T.H. Davies, A.J.Lidsey, 
	\address{Dept. of Physics \& Astronomy,
	University of Glasgow, 
	Glasgow G12 8QQ, U.K., UKQCD collaboration}
         K.Hornbostel,
         \address{Southern Methodist University,
         Dallas, TX 75252}
        G. P. Lepage,
        \address{Newman Laboratory of Nuclear Studies,
          Cornell University, Ithaca, NY 14853}
        J. Shigemitsu, 
        \address{Physics Department, 
        The Ohio State University, 
        Columbus, Ohio 43210, USA.}
        J. Sloan
        \address{Florida State University, SCRI, Tallahassee, FL 32306}
      } 
\begin{document}

\begin{abstract}

We update our results for the heavy-heavy spectrum at $\beta$ = 5.7,
using NRQCD. This includes a scaling comparison with the $\Upsilon$ 
spectrum at $\beta$ = 6.0 and first lattice predictions for the 
$B_c$ spectrum. 

\end{abstract}

\maketitle

\section{Introduction}

We have recently given accurate results for bottomonium and charmonium 
spectroscopy on the lattice \cite{ups,psi}, exploiting the non-relativistic
nature of these systems by using a non-relativistic effective theory, 
NRQCD\cite{nrqcd}. This can be 
systematically matched to full QCD, order by order
in $\alpha_s$ and $v^{2}/c^{2}$, where $v$ is the velocity of the 
heavy quark in the bound state. A limited number of terms in the 
NRQCD action suffice provided that $v^{2}$ is small and the lattice 
heavy quark mass, $Ma$ is not too small. The latter condition prevents 
$a$ being taken to zero in this effective theory, but it should still
be true that physical results scale in a limited region of $a$. This 
is all that is necessary to obtain continuum values. We test the scaling 
of our $\Upsilon$ spectrum results by comparing results at $\beta$ =5.7
with those at $\beta$ = 6.0, an updated picture of which is presented here
in the talks by Paul McCallum and Junko Shigemitsu. 

There is much current interest in the spectrum of mixed bound states of bottom
and charm quarks with the hope that they will soon be observed
experimentally. 
Here we present results for
$B_c$ spectroscopy from NRQCD and use our $\Upsilon$ and $\Psi$
results to give good indications of the size and direction of systematic
 errors. We compare our results to recent 
calculations in potential models \cite{eichten}.

\section{Lattice NRQCD}

The NRQCD action, details of how heavy quark propagators are evaluated,
the smearing functions and the fitting procedures used are described in
detail in several publications \cite{cornell,ups,psi} and are
reviewed in this Proceedings
 in the talks by Paul McCallum and Junko Shigemitsu. 

The starting point of NRQCD is to expand the original QCD 
Lagrangian in powers of $v^{2}$, the typical quark velocity in 
a bound state.
For the $J/\Psi$ system $v^{2} \sim 0.3$ and for $\Upsilon$ 
$v^{2} \sim 0.1$. For $B_c$ $v^{2} \sim 0.15$ for the single
particle with mass equal to the reduced mass of the $b$, $c$ system,
but the kinetic energy will be shared unequally between the $b$ and
$c$. $v^{2} \sim 0.5$ for the $c$ quark so relativistic corrections will
be {\it more} important than for $c\overline{c}$ systems. NRQCD enables these 
corrections to be added systematically. 

The action used here is 
the same one used in refs. \cite{ups} and \cite{psi}. Relativistic corrections
 ${\cal O} (M_Qv^{4})$ have been included for both $b$ and $c$ quarks. 
Other sources of systematic error are
discretisation errors and errors from
the absence of virtual quark loops 
because we use
quenched configurations generated with the standard
plaquette action. Finite volume errors
should be negligible because of the relatively small size of heavy-heavy
 systems.

To calculate masses for $b\overline{c}$ bound states we define 
$b$ and $c$ quark Green functions on the lattice, in the standard way. 
We used 200 quenched gluon field configurations
at $\beta = 6/g^2$ = 5.7 generously supplied by
the UKQCD collaboration \cite{thanks_ukqcd}, and fixed to Coulomb gauge, 
using a Fourier accelerated steepest descents algorithm \cite{facc}.
 The bare masses of the $c$ and 
$b$ quarks were chosen from fits to the kinetic mass of the 
$\eta_c$\cite{psi} and the $\Upsilon$, giving $am_c$ = 0.8 (with
stability parameter, $n$ = 4) and $am_b$ = 3.15 ($n$ = 2).

We combine a $b$ propagator with a $\overline{c}$ propagator to 
make $B_c$ states. From the evolution equations, it is more efficient to
have numerous $b$ quark propagators with small values of the stability 
parameter $n$ and only one $c$ quark propagator. We therefore fixed the 
$c$ quark with a $\delta$ function source and used various smearing functions 
as sources for the $b$ quark. This enabled us to also map out the 
$\Upsilon$ spectrum at the same time. 

Local meson operators are tabulated in \cite{ups}. 
Using the notation $^{2S+1}L_{J}$, we have looked at meson 
propagators for the following states: $^{1}S_{0}$, $^{3}S_{1}$, $^{1}P_{1}$,
$^{3}P_{0}$, $^{3}P_{1}$ and $^{3}P_{2}$ for both the
E and T representation.
Since $C$ is not a good quantum number for the $B_c$ system the 
$^{1}P_1$ and $^{3}P_1$ mesons will mix and so in addition we
calculated the cross-correlation between these two.
For the 
$s$ states, smearing functions both for the ground and first radially excited
state were used as well as a local $\delta$ function. {}From
this all possible combinations of smearing at the source with a 
local sink were formed.
For the $p$ states only the ground state
smearing function was used at the source.
We calculated the dispersion relation for the $\Upsilon$ and for the ($B_c$) 
by looking at the meson propagator
for small non-zero momenta. 
Because of the relatively small size of these systems it is possible to
use more than one starting site for the mesons. We used 8 different
spatial origins and 2 different starting times to increase our statistics,
but we bin correlation functions over each configuration. 

\section{Simulation results}

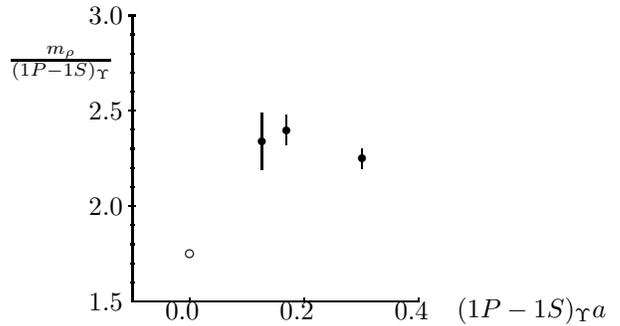
\begin{figure}
\begin{center}
\setlength{\unitlength}{.01in}
\begin{picture}(150,150)(-30,150)
\put(-30,150){\line(0,1){150}}
\put(-30,150){\line(1,0){150}}
\multiput(-32,200)(0,50){3}{\line(1,0){4}}
\multiput(-31,160)(0,10){15}{\line(1,0){2}}
\multiput(0,150)(60,0){3}{\line(0,1){2}}
\multiput(-15,150)(15,0){6}{\line(1,0){1}}
\put(-35,150){\makebox(0,0)[r]{1.5}}
\put(-35,200){\makebox(0,0)[r]{2.0}}
\put(-35,250){\makebox(0,0)[r]{2.5}}
\put(-35,300){\makebox(0,0)[r]{3.0}}
\put(-95,275){\makebox(0,0)[l]{$\frac {m_{\rho}} {(1P-1S)_{\Upsilon}}$}}
\put(5,145){\makebox(0,0)[r]{0.0}}
\put(65,145){\makebox(0,0)[r]{0.2}}
\put(125,145){\makebox(0,0)[r]{0.4}}
\put(140,145){\makebox(0,0)[l]{$\small{(1P-1S)_{\Upsilon}a}$}}

\put(90.3,225){\circle*{4}}
\put(90.3,220){\line(0,1){10}}
\put(50.7,240){\circle*{4}}
\put(50.7,232){\line(0,1){16}}
\put(37.8,234){\circle*{4}}
\put(37.8,219){\line(0,1){30}}

\put(0,175){\circle{4}}
\end{picture}
\end{center}
\caption{The scaling plot for the ratio $m_{\rho}/(1P-1S)_{\Upsilon}$ 
at 3 different values of $\beta$, filled 
circles. The $m_{\rho}$ values are preliminary UKQCD results
using the tadpole-improved clover action. The $\Upsilon$ results
at large $a$ are NRQCD results at $\beta$ = 5.7 and 6.0. The 
point at smallest $a$ is taken from a UK(NR)QCD result using only the
leading order NRQCD action. The open circle shows the experimental 
result.} 
\end{figure}

The most accurate quantities we calculate are spin-independent splittings such
as the spin-averaged 1P-1S splitting. A comparison between this splitting at 
$\beta$ = 5.7 and $\beta$=6.0 shows good scaling behaviour {\it once} $\cal{O}$$(a^2)$
errors in the gluon field configurations are removed by estimating their effect
on 1S levels either perturbatively or non-perturbatively  \cite{corrns}
(a 10\% effect for the $\Upsilon$ at $\beta$ =5.7).
Figure 1 shows a scaling plot for the ratio of the 1P-1S splitting  to $m_{\rho}$.
To see scaling we must use results for $m_{\rho}$ which have $\cal{O}$$(a)$ errors
removed. The results in Figure 1 come from UKQCD \cite{hps} using a tadpole-improved
clover action for light propagators. The $\Upsilon$ splitting at $\beta$ = 6.2. is
from ref. \cite{cambridge} using a lowest order NRQCD action only and therefore
has a potential 10\% systematic uncertainty.  

The difference with experiment for this ratio is similar to that we find 
when comparing splittings for $b\overline{b}$ and $c\overline{c}$ 
at $\beta$=5.7 \cite{psi} and interpret it as a result of incorrect running of the coupling constant
between scales in the quenched approximation. The light hadron spectrum may still
have large systematic errors arising from other sources such as finite volume so 
it is too early to put this result down entirely to quenching. 

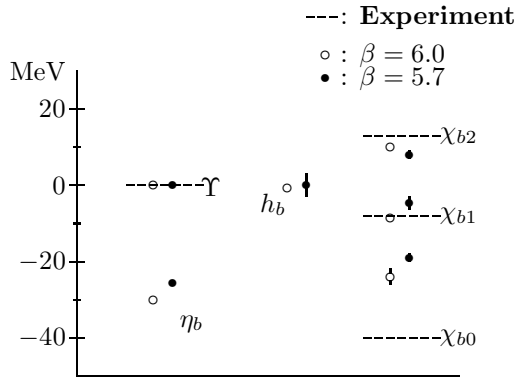
\begin{figure}
\begin{center}
\setlength{\unitlength}{.02in}
\begin{picture}(100,40)(15,-50)
\put(15,-50){\line(0,1){80}}
\multiput(13,-40)(0,20){4}{\line(1,0){4}}
\multiput(14,-40)(0,10){7}{\line(1,0){2}}
\put(12,-40){\makebox(0,0)[r]{$-40$}}
\put(12,-20){\makebox(0,0)[r]{$-20$}}
\put(12,0){\makebox(0,0)[r]{$0$}}
\put(12,20){\makebox(0,0)[r]{$20$}}
\put(12,30){\makebox(0,0)[r]{MeV}}
\put(15,-50){\line(1,0){100}}

\multiput(75,44)(3,0){3}{\line(1,0){2}}
\put(84,44){\makebox(0,0)[l]{: {\bf Experiment}}}
\put(80,34){\makebox(0,0)[tl]{\circle{2}}}
\put(84,34){\makebox(0,0)[l]{: $\beta = 6.0$}}
\put(80,28){\makebox(0,0)[tl]{\circle*{2}}}
\put(84,28){\makebox(0,0)[l]{: $\beta = 5.7$ }}

\multiput(28,0)(3,0){7}{\line(1,0){2}}
\put(50,2){\makebox(0,0)[t]{$\Upsilon$}}
\put(35,0){\circle{2}}
\put(40,0){\circle*{2}}

\put(45,-34){\makebox(0,0)[t]{$\eta_b$}}
\put(35,-29.9){\circle{2}}
\put(40,-25.6){\circle*{2}}

\put(63,-5){\makebox(0,0)[l]{$h_b$}}
\put(70,-0.8){\circle{2}}
\put(75,-0.0){\circle*{2}}
\put(75,-0.0){\line(0,1){3.0}}
\put(75,-0.0){\line(0,-1){3.0}}

\multiput(90,-40)(3,0){7}{\line(1,0){2}}
\put(110,-40){\makebox(0,0)[l]{$\chi_{b0}$}}
\put(97,-24){\circle{2}}
\put(97,-23){\line(0,1){1}}
\put(97,-25){\line(0,-1){1}}
\put(102,-19){\circle*{2}}
\put(102,-19){\line(0,1){1}}
\put(102,-19){\line(0,-1){1}}

\multiput(90,-8)(3,0){7}{\line(1,0){2}}
\put(110,-8){\makebox(0,0)[l]{$\chi_{b1}$}}
\put(97,-8.6){\circle{2}}
\put(102,-4.7){\circle*{2}}
\put(102,-4.7){\line(0,1){1.7}}
\put(102,-4.7){\line(0,-1){1.7}}

\multiput(90,13)(3,0){7}{\line(1,0){2}}
\put(110,13){\makebox(0,0)[l]{$\chi_{b2}$}}
\put(97,10.1){\circle{2}}
\put(102,8.0){\circle*{2}}
\put(102,8.0){\line(0,1){1.0}}
\put(102,8.0){\line(0,-1){1.0}}

\end{picture}
\end{center}
\caption{ A comparison of $\Upsilon$ spin splittings 
for $\beta$=5.7 and $\beta$ = 6.0. The spin-average of $s$ and
$p$ states is set to zero.}

\end{figure}

We currently determine  spin splittings only to leading order with no
discretisation corrections. This has a large effect at $\beta$ = 5.7 
as can be seen from Figure 2. The discretisation corrections tend to
reduce the overall size of splittings and also tend to equalise the
$^3P_2 - ^3P_1$ and $^3P_1 - ^3P_0$ splittings. This is already 
evident at $\beta$ = 6.0 and is worse at $\beta$ = 5.7. This can be
improved by systematically adding discretisation corrections to the 
$E$ and $B$ fields. 

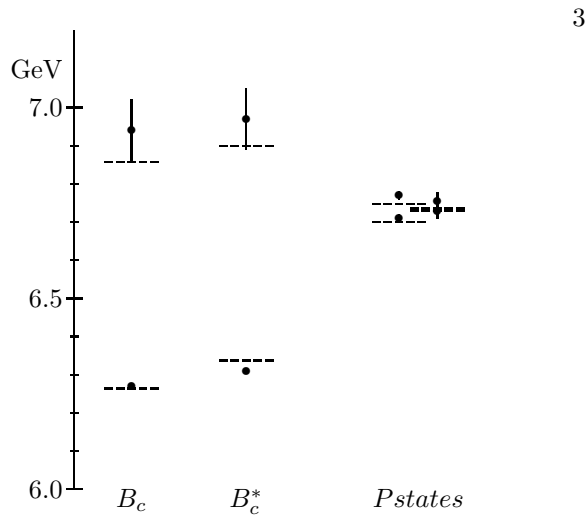
\begin{figure}
\begin{center}
\setlength{\unitlength}{.02in}
\begin{picture}(130,120)(10,580)
\put(15,600){\line(0,1){120}}
\multiput(13,600)(0,50){3}{\line(1,0){4}}
\multiput(14,600)(0,10){11}{\line(1,0){2}}
\put(12,600){\makebox(0,0)[r]{6.0}}
\put(12,650){\makebox(0,0)[r]{6.5}}
\put(12,700){\makebox(0,0)[r]{7.0}}
\put(12,710){\makebox(0,0)[r]{GeV}}

     \put(30,600){\makebox(0,0)[t]{$B_c$}}
     \put(30,627){\circle*{2}}
     \multiput(23,626.4)(3,0){5}{\line(1,0){2}}

     \put(30,694){\circle*{2}}
     \put(30,694){\line(0,1){8}}
     \put(30,694){\line(0,-1){8}}
     \multiput(23,685.6)(3,0){5}{\line(1,0){2}}

     \put(60,600){\makebox(0,0)[t]{$B^{*}_c$}}
     \put(60,631){\circle*{2}}
     \multiput(53,633.7)(3,0){5}{\line(1,0){2}}

     \put(60,697){\circle*{2}}
     \put(60,697){\line(0,1){8}}
     \put(60,697){\line(0,-1){8}}
     \multiput(53,689.9)(3,0){5}{\line(1,0){2}}

     \put(105,600){\makebox(0,0)[t]{$P states$}}
     \put(100,677){\circle*{2}}
     \put(100,677){\line(0,1){1}}
     \put(100,677){\line(0,-1){1}}
     \multiput(93,674.7)(3,0){5}{\line(1,0){2}}

     \put(100,671){\circle*{2}}
     \put(100,671){\line(0,1){1}}
     \put(100,671){\line(0,-1){1}}
     \multiput(93,670.0)(3,0){5}{\line(1,0){2}}

     \put(110,673){\circle*{2}}
     \put(110,673){\line(0,1){2}}
     \put(110,673){\line(0,-1){2}}
     \put(110,675.6){\circle*{2}}
     \put(110,675.6){\line(0,1){2}}
     \put(110,675.6){\line(0,-1){2}}
     \multiput(103,673.0)(3,0){5}{\line(1,0){2}}
     \multiput(103,673.6)(3,0){5}{\line(1,0){2}}


\end{picture}
\end{center}
\caption{NRQCD simulation results for the spectrum of the
$B_c$ system 
using an inverse lattice spacing of 1.32~GeV.
Error bars are shown where visible and only indicate statistical
uncertainties.}
\label{fig:bc}
\end{figure}

Figure 3 shows results for the $B_c$ spectrum. We are able to  
predict the overall mass scale as well as the splittings since the 
$c$ and $b$ masses were fixed from the $\Upsilon$ and $\eta_c$ masses.
The fact that $a^{-1}$ is different for $c\overline{c}$ and $b\overline{b}$ 
in the quenched approximation means that we must fix $a^{-1}$ for $b\overline{c}$
separately. We do this by setting the 1P-1S splitting to be the same as that 
for $b\overline{b}$ and $c\overline{c}$ (it is $M_Q$ independent). This gives
an $a^{-1}$ between the two previous values and means that the quark masses in 
physical units are not the same in the three systems. This is an unavoidable 
artefact of quenching and adds to the systematic error. 

The $B_c$ mass is obtained by calculating correlation functions at non-zero
momenta to derive the dispersion relation. The mass is the kinetic mass in the
equation:
\begin{equation}
E(p) = E_{0} + p^2/2M_{kin} .
\end{equation}  
$M_{kin}$ will not agree with the static mass, $E_0$, because the quark mass
has been removed from the Hamiltonian. There is then a shift
between $E_0$ and $M_{kin}$ that must be applied per quark in a meson. 
We can check non-perturbatively that the shift for the $B_c$ is the 
average of that for $c\overline{c}$ and $b\overline{b}$. This is shown in 
Table 1 and works very well. For the clover action at large mass it does
not work, and instead the shift is dependent, incorrectly, on the
 meson itself \cite{sarajohn}.  

\begin{table}
\begin{center}
\begin{tabular}{c|lll}
 & $E_0$ & $M_{kin}$ & shift  \\
\cline{1-4}
$\Upsilon$  & 0.5030  & 7.04 & 6.54  \\
$\eta_c$  & 0.618 & 2.430 & 1.812 \\
$B_c$ & 0.6052 & 4.79 & 4.18 \\
\cline{1-4}
\end{tabular}
\end{center}
\caption{ Results for static and kinetic masses and the difference
between them for heavy-heavy mesons, using NRQCD at \protect $\beta$
=5.7. Values are in lattice units. 
}
\label{shifts}
\end{table}

The comparison of our results with a potential model 
shows fairly good agreement (Figure 3) and indicates that
relativistic effects may fortuitously not be very large.
 We have included the first relativistic corrections but potential
model calculations include relativistic effects only through the
phenomenological potential. Better estimates of these corrections need
to be made.

Our experience with
heavy-heavy spectra in the quenched approximation indicates that our calculation is  
likely to overestimate the 2S-1S splitting and to underestimate the hyperfine and 
$p$ states splittings. Discretisation errors also reduce our spin splittings, as above. 
Unquenching and discretisation corrections would then bring us 
into better agreement with the 
potential model results {\it except} for the $p$ fine structure 
where we already have larger
splittings. We believe as a result that the potential model calculation is 
underestimating these splittings. 

Our preliminary value for the $B_c$ decay constant is 460 MeV. This 
calculation includes only the leading order term in the heavy-light current.
There is a $1/M_c$ correction for $B_c$ whereas in $b\overline{b}$ and
$c\overline{c}$ the first correction appears at $1/M_Q^2$. 

After this talk was given we learned that the ALEPH collaboration has 
candidate $B_c$ events \cite{aleph}.

\section{Summary}

We present heavyonium spectrum results at $\beta$ = 5.7, including first
lattice predictions for
the $B_c$ spectrum. Future calculations will include further relativistic
and discretisation corrections as well as a comparison to unquenched 
results at similar lattice spacing. 

\vspace{.1in}
\noindent
This work is supported in part by the U.S. DOE, the NSF and by the UK PPARC. 
The numerical computations were carried out at NERSC 
and the Atlas Centre. We thank UKQCD
for providing the quenched configurations

\end{document}